\documentclass[aps,prd,eqsecnum,12pt]{revtex4-1}
\usepackage{epsfig}
\newcommand{\beq}{\begin{equation}}
\newcommand{\eeq}{\end{equation}}
\newcommand{\beqa}{\begin{eqnarray}}
\newcommand{\eeqa}{\end{eqnarray}}
\newcommand{\lslash}[1]{#1\llap/}

\newcommand{\Eq}[1]{Eq.\ (\ref{#1})}
\newcommand{\Eqs}[2]{Eqs.\ (\ref{#1}) and (\ref{#2})}
\newcommand{\Eqss}[3]{Eqs.\ (\ref{#1}), (\ref{#2}) and (\ref{#3})}
\newcommand{\Eqsss}[4]{Eqs.\ (\ref{#1}), (\ref{#2}), (\ref{#3}) and (\ref{#4})}
\newcommand{\Ref}[1]{Ref.\ \cite{#1}}
\newcommand{\Section}[1]{Section\ \ref{#1}}
\newcommand{\Fig}[1]{Fig.\ \ref{#1}}
\newcommand{\Tr}{\mbox{Tr}\,}
\newcommand{\gtilde}{\tilde g}

\begin{document}

\title{Chiral spin-3/2 particles in a medium}

\author{Jos\'e F. Nieves}
\email{nieves@ltp.uprrp.edu}
\affiliation{Laboratory of Theoretical Physics, Department of Physics\\
  University of Puerto Rico, R\'{\i}o Piedras, Puerto Rico 00936
}

\author{Sarira Sahu}
\email{sarira@nucleares.unam.mx}
\affiliation{Instituto de Ciencias Nucleares, Universidad Nacional
  Aut\'onoma de Mexico,\\
  Circuito Exterior, C. U., A. Postal 70-543, 04510 Mexico DF, Mexico
}

\date{March 2016}

\begin{abstract}
We consider the propagation of a chiral spin-3/2 particle
in a background medium using the Thermal Field Theory (TFT) method,
in analogy to the cases of a spin-1/2 fermion (e.g., a neutrino)
and the photon. We present a systematic decomposition of the
thermal self-energy, from which the dispersion relation of
the modes that propagate in the medium are obtained.
We find that there are several modes and 
in each case we obtain the equation for the dispersion relation
as well as the corresponding spin-3/2 spinor.
As an example of the general procedure and results,
we consider a model in which the chiral spin-3/2
particle couples to a spin-1/2 fermion and a scalar particle,
and propagates in a thermal background composed of such particles.
The dispersion relations and corresponding spinors are detemined
explcitly in this case from the 1-loop TFT expression for the self-energy.
The results in this case share some
resemblance and analogies with the photon and the chiral fermion cases but,
as already noted, there are also differences. The present work provides
the groundwork for considering problems related to the properties
of chiral spin-3/2 particles in a medium, in analogy to the case
of neutrinos for example, which can be relevant in physical
contexts of current interest.
\end{abstract}
\maketitle

\section{Introduction}

It is known that dispersion effects can
have significant impact on the properties of elementary particles
when they propagate through a background medium, such as the
dispersion relation and the induced electromagnetic couplings of electrically
neutral particles. The effects on photons and plasma physics
have of course been known for a long time, and more recently it has
been a crucial item in neutrino physics since the discovery of the
MSW effect\cite{wolfenstein,*ms,*langacker}.
From a modern point of view, the methods of Thermal Field Theory
(TFT)\cite{*[{See for example,}] [] landsman,*kapusta,*bellac}
have been useful for studying many problems associated with such effects
in a variety of physical contexts. This view has been partially
stimulated by the original work of Weldon\cite{weldon:fermions,weldon:cov}
showing the convenience of the covariant TFT calculations
in this kind of problems, in particular in systems involving
chiral fermions at finite temperature.

In the case of neutrinos there is an extensive literature
on the effects of the background medium on their properties
and propagation. Apart from the dispersion relation\cite{%
notzold-raffelt,*palpham,jfn1,*dnt}, the medium also induces electromagnetic
couplings\cite{*[{See, }] [{ and references therein}] raffelt:book} 
that can lead to effects in astrophysical
and/or cosmological settings, as well as neutrino collective
oscillations\cite{nunu1,*nunu2,*nunu3} that have been the subject
and significant work in the context
of instabilities in supernovas\cite{*[{See for example, }]
[{, and references therein}] nunu4}.

Here we consider the propagation of a
chiral spin-3/2 particle $\lambda^\mu_{L}$ in a background medium,
using the same TFT techniques. It has been shown recently\cite{adler1,*adler2}
that the theory of a gauged massless chiral Rarita-Schwinger field\cite{das}
is consistent with physical principles (e.g., no superluminal propagation
and others).  Thus it seems useful to look in a general way at the
case of a massless chiral spin-3/2 particle propagating in a thermal
background, in analogy to the case of a spin-1/2 fermion (e.g., a neutrino)
or the photon, as mentioned above. Our aim is to provide
a useful starting point for considering similar problems in the spin-3/2 case.
Because the chiral $\lambda^\mu_{L}$ field can be thought of as a combination
of a spin-1/2 chiral field and massless spin-1 field, the problem shares
some analogies with both of those cases, although the details
are different. 

Our main result is a systematic decomposition of the
thermal self-energy, from which the dispersion relation of
the modes that propagate in the medium are obtained.
We assume that the interactions of the $\lambda_L$ particle
with the thermal background particles are such that the thermal self-energy is
transverse to the momentum four-vector of the propagating mode.
A key element of the procedure is a general expression
for the self-energy in terms of a set of scalar functions,
each one corresponding to an independent tensor, consistent with
the transversality condition, constructed using
the momentum vectors available in the system (the momentum four-vector $k^\mu$
of the particle and the background velocity four-vector $u^\mu$), as well
as the gamma matrices, the metric tensor and the Levi-Civita tensor.
On the basis of such a decomposition we find that there is a
transverse mode, in the sense that its spin-3/2 spinor is transverse
to $k^\mu$, and two other modes that involve the longitudinal polarization
vector. In each case we obtain the equation for the dispersion relation
in terms of the self-energy scalar functions, as well as the corresponding
spin-3/2 spinor. We illustrate the application of the procedure and
the results by considering a model in which the $\lambda^\mu_L$
particle couples to a spin-1/2 fermion and a scalar particle,
and propagates in a thermal background composed of such particles.
From the 1-loop TFT expression for the self-energy, we determine
the scalar functions referred to above, and in turn from them
the dispersion relations and corresponding spinors.
The results in this case share some
resemblance and analogies with the photon and the chiral fermion case but,
as already noted, there are also differences. Thus, the present work provides
the groundwork for considering problems related to the properties
of chiral spin-3/2 particles in a medium, in analogy to the case
of neutrinos for example.
The results presented here can be useful in problems of current research
interest that involve the thermal production of
spin-3/2 particles in cosmological (e.g., gravitinos in the Early Universe)
or astrophysical contexts\cite{cosmology1,*cosmology2,*cosmology3,*cosmology4,%
darkmatter1,*darkmatter2,*darkmatter3}.

The rest of the paper is organized as follows. In \Section{sec:selfenergy} we
introduce the self-energy function and the $\lambda^\mu_L$
effective equation of motion in a medium. In \Section{sec:generalform}
we write down the decomposition of the thermal self-energy in terms of a set of
independent structure tensors and the corresponding scalar functions,
consistent with the transversality condition.
The tensors are constructed from the momentum of the particle
$k^\mu$ and the background velocity four-vector $u^\mu$, as well
as the gamma matrices, the metric tensor and the Levi-Civita tensor.
In \Section{sec:basisspinors} we set down the conventions that we use
for the spin-3/2 basis spinors, in terms of the spin-1 polarization
vectors and the spin-1/2 chiral spinors. In \Section{sec:disprel}
the equations for the dispersion relations are obtained, and
finally in \Section{sec:model} the equations are solved explicitly
in the example mentioned above.

\section{Equation of motion and the self-energy in the medium}
\label{sec:selfenergy}

The free-field part of the chiral (massless) RS Lagrangian in coordinate space
is given by
\beq
\label{LRSvacuum}
L^{(0)}_\lambda = -\bar\lambda^\mu_L\left\{i\lslash{\partial}\lambda_{L\mu}
- \gamma_\mu i\partial\cdot\lambda_L - i\partial_\mu\gamma\cdot\lambda_L
+ \gamma_\mu i\lslash{\partial}\gamma\cdot\lambda_L\right\}\,,
\eeq
and in momentum space it translates to
\beq
\label{Lkvacuum}
L^{(0)}_\lambda(k) = \bar\lambda^\mu_L(k) L_{\mu\nu} \lambda^\nu_L(k) \,,
\eeq
where
\beq
\label{Lmunu1}
L_{\mu\nu} = -\left[g_{\mu\nu}\lslash{k} - k_\mu\gamma_\nu
- k_\nu\gamma_\mu + \gamma_\mu\lslash{k}\gamma_\nu\right]\,.
\eeq
\Eqs{LRSvacuum}{Lkvacuum} can be rewritten in the alternative forms
\beq
\label{L0lambdaalternative}
L^{(0)}_\lambda = -\epsilon_{\mu\nu\alpha\beta}
\bar\lambda^\mu_L\gamma^\alpha\partial^\beta\lambda^\nu_L\,,
\eeq
and
\beq
L^{(0)}_\lambda(k) = \bar\lambda^\mu_L(k) \ell_{\mu\nu\lambda}(k)\gamma^\lambda
\lambda^\nu_L(k) \,,
\eeq
respectively, where we have defined
\beq
\label{elldef}
\ell_{\mu\nu\lambda}(k) = i\epsilon_{\mu\nu\lambda\rho} k^\rho \,.
\eeq
This follows from the identity\footnote{We use the convention
$\gamma_5 = i\gamma^0 \gamma^1 \gamma^2 \gamma^3$ and
$\epsilon^{0123} = +1$.}
\beq
\label{threegammaidentity}
\gamma_\alpha\gamma_\beta\gamma_\gamma =
(g_{\alpha\beta}\gamma_\gamma - g_{\alpha\gamma}\gamma_\beta
+ g_{\beta\gamma}\gamma_\alpha) +
i\epsilon_{\alpha\beta\gamma\lambda} \gamma^\lambda\gamma_5\,,
\eeq
which implies, for example, that
\beq
\label{ellLrel}
L_{\mu\nu}L = \ell_{\mu\nu\lambda}\gamma^\lambda L\,,
\eeq
where we have defined as usual $L = \frac{1}{2}(1 - \gamma_5)$.

Chirality implies that, in the presence of the medium,
the effective Lagrangian has a similar gamma matrix structure.
In particular it can be written in the form
\beq
\label{pimunulambdadef}
L_{eff}(k) = \bar\lambda^\mu_L(k)
\left[\ell_{\mu\nu\lambda} - \pi_{\mu\nu\lambda}\right]
\gamma^\lambda\lambda^\nu_L(k) \,,
\eeq
where $\pi_{\mu\nu\lambda}$ is a tensor that depends on
$k^\mu$ and the velocity four-vector of the background medium $u^\mu$,
but does not contain any $\gamma$ matrices. Therefore the
thermal self-energy $\Sigma_{T\mu\nu}$ must have the form
\beq
\label{Sigmapirel}
\Sigma_{T\mu\nu} = \pi_{\mu\nu\lambda}\gamma^\lambda L\,,
\eeq
which in turn implies the convenient formula
\beq
\label{pisigmarel}
\pi_{\mu\nu\lambda} = \frac{1}{2}\Tr\left(
\gamma_\lambda\Sigma_{T\mu\nu}\right)\,.
\eeq
The dispersion relations of the propagating modes
then follow from solving the effective RS equation in the medium,
\beq
\label{effeqmotion1}
\left[\ell_{\mu\nu\lambda} -
\pi_{\mu\nu\lambda}\right]\gamma^\lambda\psi^\nu_L(k) = 0\,,
\eeq
or equivalently
\beq
\label{effeqmotion2}
-\left[g_{\mu\nu}\lslash{k} - k_\mu\gamma_\nu
- k_\nu\gamma_\mu + \gamma_\mu\lslash{k}\gamma_\nu\right]\psi^\nu_L(k)
- \pi_{\mu\nu\lambda}\gamma^\lambda\psi^\nu_L(k) = 0\,.
\eeq
Our purpose in what follows is to use this equation as the starting point
to calculate the dispersion relation of the $\lambda$ propagating modes
in a background medium as previously described.

\section{General form of $\pi_{\mu\nu\alpha}$}
\label{sec:generalform}

The tensor $\pi_{\mu\nu\lambda}$ introduced in \Eq{pimunulambdadef}
depends on the vectors $k^\mu$ and $u^\mu$, but does not contain any
$\gamma$ matrices, and since we are assuming that the interactions
of the $\lambda_L$ particle are such that the thermal self-energy is
transverse, it satisfies the transversality condition,
\beq
k^\mu \pi_{\mu\nu\alpha} = k^\nu \pi_{\mu\nu\alpha} = 0\,.
\eeq
In analogy with the decomposition of the photon self-energy in a medium,
it is also convenient here to introduce a similar notation. Thus we define
\beq
\label{gtilde}
\gtilde_{\mu\nu} = g_{\mu\nu} - \frac{k_\mu k_\nu}{k^2}\,,
\eeq
the transverse vector
\beq
\label{utilde}
\tilde u_\mu = \gtilde_{\mu\nu} u^\nu \,,
\eeq
and the tensors
\beqa
\label{RQP}
R_{\mu\nu} & = & \gtilde_{\mu\nu} - Q_{\mu\nu} \,,\nonumber\\
Q_{\mu\nu} & = & \frac{\tilde u_\mu\tilde u_\nu}{\tilde u^2}\,,\nonumber\\
P_{\mu\nu} & = & \frac{i}{\kappa}\epsilon_{\mu\nu\alpha\beta}k^\alpha u^\beta
\,,
\eeqa
where
\beq
\label{kappa}
\kappa = \sqrt{\omega^2 - k^2} \,,
\eeq
with
\beq
\label{omega}
\omega = k\cdot u\,.
\eeq
The variables $\omega$ and $\kappa$ are simply the energy and magnitude
of the three-dimensional momentum vector, respectively, in the frame in
which the medium is at rest, i.e., the frame in which $u^\mu = (1, 0)$.

With this notation at hand, it is clear that, aside from a term
$\sim \epsilon_{\mu\nu\alpha\beta}k^\beta$ of similar
form to the vacuum kinetic energy term, $\pi_{\mu\nu\alpha}$ can be
written as a combination of terms of the following form
\beq
T_{\mu\nu} a_\alpha\;, T_{\mu\alpha}\tilde u_\nu\;,
T_{\alpha\nu}\tilde u_\mu\;, \epsilon_{\mu\nu\alpha\beta}k^\beta\,,
\eeq
where $T$ can be either $R$, $Q$ or $P$, defined in \Eq{RQP},
and $a_\alpha$ can be either $k_\alpha$ or $u_\alpha$. Therefore we write
\beq
\label{piprimedef}
\pi_{\mu\nu\alpha} = \pi_0 \ell_{\mu\nu\alpha} + \pi^\prime_{\mu\nu\alpha}\,,
\eeq
with
\beqa
\label{piparametrization}
\pi^\prime_{\mu\nu\alpha} & = & \pi_{R1} R_{\mu\nu} k_\alpha +
\pi_{R2} R_{\mu\nu} u_\alpha +
\pi_{R3} R_{\mu\alpha} \tilde u_\nu +
\pi_{R4} R_{\alpha\nu} \tilde u_\mu\nonumber\\
&&\mbox{} +
\pi_{P1} P_{\mu\nu} k_\alpha +
\pi_{P2} P_{\mu\nu} u_\alpha +
\pi_{P3} P_{\mu\alpha} \tilde u_\nu +
\pi_{P4} P_{\alpha\nu} \tilde u_\mu\nonumber\\
&&\mbox{} +
\pi_{Q1} Q_{\mu\nu} k_\alpha +
\pi_{Q2} Q_{\mu\nu} u_\alpha\,,
\eeqa
where we have used the fact that the terms $Q_{\mu\alpha}\tilde u_\nu$ and
$Q_{\alpha\nu}\tilde u_\mu$ are of the same form as those as the
$\pi_{Q1,2}$ terms above, and $\ell_{\mu\nu\alpha}$
is defined in \Eq{elldef}.

\section{Basis spinors}
\label{sec:basisspinors}

\subsection{spin-1 polarization vectors}

It is useful to introduce the following notation, borrowed from
the analogous discussions in the case of the photon propagating in a medium.
Adopting the rest-frame of the medium,
\beq
u^\mu = (1,\vec 0) \,,
\eeq
and writing the momentum vector in the form
\beq
k^\mu = (\omega, \vec k)\,,
\eeq
we introduce the spin-1 polarization vectors in the usual way,
\beqa
\epsilon^\mu_{1,2} & = & (0, \hat e_{1,2}) \,,\nonumber\\
\epsilon^\mu_\pm & = & \frac{1}{\sqrt{2}}\left(
\epsilon^\mu_{1} \pm i\epsilon^\mu_{2}\right)\,,
\eeqa
where $\hat e_{1,2}$ are such that
\beq
\hat e_{1,2}\cdot\vec k = 0\,,\qquad
\hat e_2 = \vec k \times \hat e_1\,.
\eeq
In addition we define
\beq
\label{epsilonell}
\epsilon^\mu_{\ell} = \frac{1}{\sqrt{-\tilde u^2}}\tilde u^\mu\,,
\eeq
where $\tilde u_\mu$ has been defined in \Eq{utilde}, which in 
the rest frame of the medium is given by
\beq
\epsilon^\mu_{\ell} = -\frac{1}{\sqrt{k^2}}(\kappa, \omega\hat k) \,,
\eeq
with $\kappa$ defined in \Eq{kappa}. These vectors are mutually
orthogonal and satisfy the following relations
\beq
\label{transversality}
\epsilon_{\ell}\cdot k = \epsilon_\pm\cdot k = \epsilon_\pm\cdot u = 0\,,
\eeq
which in turn imply
\beqa
\label{RQPepsilon}
%{R^\mu}_{\nu}\epsilon^\nu_\pm & = & \epsilon^\mu_\pm\,, \nonumber\\
R_{\mu\nu}\epsilon^\nu_\pm & = & \epsilon_{\pm\mu}\,, \nonumber\\
%{Q^\mu}_{\nu}\epsilon^\nu_{\ell} & = & \epsilon^\mu_{\ell}\,,\nonumber\\
Q_{\mu\nu}\epsilon^\nu_{\ell} & = & \epsilon_{\ell\mu}\,,\nonumber\\
%{P^\mu}_{\nu}\epsilon^\nu_{\pm} & = & \pm\epsilon^\mu_{\pm}\,,\nonumber\\
P_{\mu\nu}\epsilon^\nu_{\pm} & = & \pm\epsilon_{\pm\mu}\,,\nonumber\\
%{R^\mu}_{\nu}\epsilon^\nu_{\ell} =
%{Q^\mu}_{\nu}\epsilon^\nu_{\pm} =
%{P^\mu}_{\nu}\epsilon^\nu_{\ell} & = & 0\,.
R_{\mu\nu}\epsilon^\nu_{\ell} =
Q_{\mu\nu}\epsilon^\nu_{\pm} =
P_{\mu\nu}\epsilon^\nu_{\ell} & = & 0\,.
\eeqa
$R^{\mu\nu}$ and $P^{\mu\nu}$ have the representation
\beqa
\label{RPalternaterep}
R^{\mu\nu} & = & -(\epsilon^\mu_{+}\epsilon^\nu_{-} +
\epsilon^\mu_{-}\epsilon^\nu_{+})\,,\nonumber\\
P^{\mu\nu} & = & -(\epsilon^\mu_{+}\epsilon^\nu_{-} -
\epsilon^\mu_{-}\epsilon^\nu_{+})\,,
\eeqa
and also useful are the relations
%
%\beqa
%\label{epsilonidentities12}
%\epsilon_{\mu\nu\alpha\beta}k^\alpha\epsilon^\beta_1 =
%\frac{\kappa}{\sqrt{-\tilde u^2}}
%\left(\epsilon_{2\mu}\epsilon_{\ell\nu} - \epsilon_{\ell\mu}\epsilon_{2\nu}
%\right)\,,\nonumber\\
%
%\epsilon_{\mu\nu\alpha\beta}k^\alpha\epsilon^\beta_2 =
%\frac{-\kappa}{\sqrt{-\tilde u^2}}
%\left(\epsilon_{1\mu}\epsilon_{\ell\nu} - \epsilon_{\ell\mu}\epsilon_{1\nu}
%\right)\,,
%\eeqa
%
%and the following ones implied by them
%
%\beq
%\label{epsilonidentities}
%\epsilon_{\mu\nu\alpha\beta}k^\alpha\epsilon^\beta_{\pm} =
%\frac{\mp i\kappa}{\sqrt{-\tilde u^2}} g_{\mu\alpha} g_{\nu\beta}
%\left(\epsilon^\alpha_{\pm}\epsilon^\beta_\ell -
%\epsilon^\alpha_\ell\epsilon^\beta_{\pm}\right)\,.
%\eeq
%
\beq
\label{epsilonidentitiesplusminus}
\epsilon_{\mu\nu\alpha\beta}k^\alpha\epsilon^\beta_{\pm} =
\frac{\mp i\kappa}{\sqrt{-\tilde u^2}}
\left(\epsilon_{\pm\mu}\epsilon_{\ell\nu} -
\epsilon_{\ell\mu}\epsilon_{\pm\nu}\right)\,.
\eeq
They can be verified as follows. Consider for example,
$\epsilon_{\mu\nu\alpha\beta}k^\alpha\epsilon^\beta_{+}$.
Since it is an antisymmetric tensor and is transverse to $k^\mu$ and
$\epsilon^\mu_{+}$, it must be proportional to the term given in the
right-hand side of \Eq{epsilonidentitiesplusminus}. The proportionality
factor can be verified by multiplying both sides with $u^\nu$ and 
reducing the resulting expressions using the multiplication rules
of the polarization vectors.

It should be kept in mind that in \Eq{epsilonell}, and wherever
$\epsilon^\mu_{\ell}$ appears, the assumption is that we are considering
situations in which $k^\mu$ is such that
\beq
k^2 \not= 0\,.
\eeq

\subsection{Dirac spinors}

Regarding the Dirac spinors we adopt, once and for all,
the Weyl representation of the gamma matrices,
\beq
\gamma^0 =
\left(\begin{array}{cc}
0 & 1\\
1 & 0
\end{array}\right)\,,\qquad
\vec\gamma =
\left(\begin{array}{cc}
0 & -\vec\sigma\\
\vec\sigma & 0
\end{array}\right)\,.
\eeq
We denote by $u_{L-}$ the left-handed chiral spinor,
\beq
\label{weylspinorL-}
u_{L-} = \left(\begin{array}{cc}
0\\
\xi_{-}
\end{array}\right)\,,
\eeq
where
$\xi_{-}$ is the Pauli spinor with negative helicity; i.e., it satisfies
\beq
\hat k\cdot\vec\sigma\, \xi_{-} = -\xi_{-} \,.
\eeq
Using the same notation, it is useful to introduce also the spinor
\beq
\label{weylspinorL+}
u_{L+} = \left(\begin{array}{cc}
0\\
\xi_{+}
\end{array}\right)\,,
\eeq
where,
\beq
\hat k\cdot\vec\sigma\, \xi_{+} = \xi_{+} \,,
\eeq
and similarly,
\beq
\label{weylspinorRpm}
u_{R\pm} = \left(\begin{array}{cc}
\xi_{\pm}\\
0
\end{array}\right)\,.
\eeq
The spinors satisfy the following relations(the right-handed counterparts
satisfy analogous relations, but we will not need them in what follows)
\beqa
\label{slashuLrelations}
\lslash{k}u_{L\pm} & = & (\omega \pm \kappa)u_{R\pm}\,,\nonumber\\
\lslash{u}u_{L\pm} & = & u_{R\pm}\,,
\eeqa
and
\beqa
\label{epsilondotgammauL}
(\epsilon_{-}\cdot\gamma) u_{L-} & = & (\epsilon_{+}\cdot\gamma) u_{L+} = 0\,
\nonumber\\
(\epsilon_{\pm}\cdot\gamma) u_{L\mp} & = & \sqrt{2}u_{R\pm}\,.
\eeqa
From \Eqs{RPalternaterep}{epsilondotgammauL} we can obtain other
useful formulas, for example
\beqa
\label{RPgammauL}
\left({R^\mu}_\alpha\gamma^\alpha\right)u_{L\pm} & = &
-\sqrt{2}\epsilon^\mu_{\pm} u^{(\mp)}_R\,,\nonumber\\
\left({P^\mu}_\alpha\gamma^\alpha\right)u_{L\pm} & = &
\mp\sqrt{2}\epsilon^\mu_{\pm} u^{(\mp)}_R\,.
\eeqa

\section{Dispersion relations}
\label{sec:disprel}

\subsection{Transverse mode}

Let us consider the spin-3/2 spinors
\beq
\label{Ut}
U^{\mu}_{L\pm} = \epsilon^\mu_{\pm} u_{L\pm} \,.
\eeq
\Eqs{transversality}{epsilondotgammauL} imply that they satisfy
\beq
\label{transversalityUt}
k\cdot U_{L\pm} = u\cdot U_{L\pm} = \gamma\cdot U_{L\pm} = 0\,,
\eeq
which in turn can be used together with \Eq{ellLrel} to obtain
\beq
\label{ellUt}
\ell_{\mu\nu\alpha}\gamma^\alpha U^\nu_{L\pm} =
-g_{\mu\nu}\lslash{k} U^\nu_{L\pm} \,.
\eeq
%
%
%
% From \Eq{RQPepsilon} we obtain the corresponding relations
%
%\beqa
%\label{RQPUt}
%{R^\mu}_\nu U^\nu_{L\pm} & = & U^\mu_{L\pm}\,, \nonumber\\
%{Q^\mu}_\nu U^\nu_{L\pm} & = & 0\,,\nonumber\\
%{P^\mu}_\nu U^\nu_{L\pm} & = & \pm U^\mu_{L\pm}\,.
%\eeqa
%
%
We now consider $\pi^\prime_{\mu\nu\alpha}$ given in \Eq{piparametrization}.
With the help of \Eqs{RQPepsilon}{epsilondotgammauL} it follows simply
\beq
(\pi^\prime_{\mu\nu\alpha}\gamma^\alpha)U^\nu_{L\pm} = g_{\mu\nu}\left[
(\pi_{R1} \pm \pi_{P1})\lslash{k} + (\pi_{R2} \pm \pi_{P2})\lslash{u}
\right]U^\nu_{L\pm}\,.
\eeq
Combining this with \Eqs{piprimedef}{ellUt} we obtain
\beq
(\pi_{\mu\nu\alpha}\gamma^\alpha)U^\nu_{L\pm} = -g_{\mu\nu}\left(
a_\pm\lslash{k} + b_\pm\lslash{u}\right)U^\nu_{L\pm}\,,
\eeq
where
\beqa
\label{abpm}
a_\pm & = & \pi_0 \mp \pi_{P1} - \pi_{R1}\,, \nonumber\\
b_\pm & = & \mp\pi_{P2} - \pi_{R2}\,.
\eeqa
Thus, if we set $\psi^\mu_L = U^\mu_{L-}$ in the equation
of motion \Eq{effeqmotion1}, using the above relations the equation becomes
\beq
\lslash{V} U^\mu_{L-} = 0\,,
\eeq
or equivalently
\beq
\label{VminusuL}
\lslash{V} u_{L-} = 0\,,
\eeq
where
\beq
V_\alpha = (1 - a_{-})k_\alpha - b_{-} u_\alpha \,.
\eeq
Recalling \Eq{slashuLrelations}, \Eq{VminusuL} requires that
$\omega$ satisfies
\beq
\label{transdisprel1}
\left[1 - a_{-}(\kappa,\omega)\right](\omega - \kappa) - b_{-}(\kappa,\omega)
 = 0\,,
\eeq
or equivalently
\beq
\label{transdisprel2}
\omega = \kappa + \frac{b_{-}(\kappa,\omega)}{1 - a_{-}(\kappa,\omega)}\,,
\eeq
where we have explicitly indicated the arguments of $a_{-}$ and $b_{-}$
to emphasize that these are implicit equations for $\omega(\kappa)$.

Another solution is $\psi^\mu_L = U^\mu_{L+}$, provided that
\beq
\lslash{V}^\prime u_{L+} = 0\,,
\eeq
where
\beq
V^\prime_\alpha = (1 - a_{+})k_\alpha - b_{+} u_\alpha \,.
\eeq
Using \Eq{slashuLrelations} as before, this yields a dispersion
relation $\omega(\kappa) = -\bar\omega(\kappa)$,
with $\bar\omega(\kappa)$ satisfying
\beq
\label{transdisprel1neg}
\left[1 - a_{+}(\kappa,-\bar\omega)\right]\bar\omega +
b_{+}(\kappa,-\bar\omega) =
\left[1 - a_{+}(\kappa,-\bar\omega)\right]\kappa\,.
\eeq
or equivalently
\beq
\label{transdisprel2neg}
\bar\omega = \kappa -
\frac{b_{+}(\kappa,-\bar\omega)}{1 - a_{+}(\kappa,-\bar\omega)}\,.
\eeq
This solution corresponds to the antiparticle propagating in the medium
with a dispersion relation $\bar\omega(\kappa)$.

These equations for the dispersion relations
(e.g., \Eqs{VminusuL}{transdisprel1}) resemble the corresponding
formulas obtained for the neutrino (or more generally a chiral spin-1/2
fermion) case\cite{weldon:fermions,jfn1,dnt}.
As we will see below, the equations involving the longitudinal
spinor are more complicated.

\subsection{Longitudinal modes}

We seek the longitudinal solution as a combination of the spinors
\beqa
\label{UL12}
U^\mu_{L1} & = & \epsilon^\mu_{\ell} u_{L-}\,,\nonumber\\
U^\mu_{L2} & = & \epsilon^\mu_{-}u_{L+}\,.
\eeqa
We thus write the solution in the form
\beq
\label{longitudinalansatz}
\psi^\mu_{L} = \sum_{a = 1,2} \alpha_a U^\mu_{La}\,,
\eeq
with coefficients $\alpha_{1,2}$ to be determined. The next step is to
derive the formulas for $\ell_{\mu\nu\alpha}\gamma^\alpha U^\nu_{L1,2}$
and $\pi_{\mu\nu\alpha}\gamma^\alpha U^\nu_{L1,2}$, the details of which
are given in the Appendix. The results given there in
\Eqsss{ellUL1}{piprimeUL1}{ellUL2}{piprimeUL2}
can be summarized in a compact form by introducing the right-handed spinors
\beqa
U^\mu_{R1} & = & \epsilon^\mu_{\ell} u_{R-}\,,\nonumber\\
U^\mu_{R2} & = & \epsilon^\mu_{-}u_{R+}\,.
\eeqa
Thus,
\beqa
\label{ellpiULell}
\left(\ell_{\mu\nu\alpha}\gamma^\alpha\right) U^\nu_{La} & = &
\sum_b L_{ba}U_{Rb\mu}\,,\nonumber\\
\left(\pi_{\mu\nu\alpha}\gamma^\alpha\right) U^\nu_{La} & = &
\sum_b \Pi_{ba}U_{Rb\mu}\,,\nonumber\\
\eeqa
where
\beqa
\label{Ldef}
L_{11} & = & 0 \,,\nonumber\\
L_{12} = L_{21}& = & \sqrt{2k^2} \,,\nonumber\\
L_{22} & = & \omega + \kappa\,,
\eeqa
and
\beqa
\label{Pi}
\Pi_{11} & = & (\omega - \kappa)\pi_{Q1} + \pi_{Q2} \,,\nonumber\\
\Pi_{12} & = & \sqrt{-2\tilde u^2}(\pi_{R4} - \pi_{P4}) +
\sqrt{2k^2}\pi_0\,,\nonumber\\
\Pi_{21} & = & \sqrt{-2\tilde u^2}(\pi_{R3} - \pi_{P3}) +
\sqrt{2k^2}\pi_0\,,\nonumber\\
\Pi_{22} & = & (\omega + \kappa)(\pi_0 + \pi_{R1} - \pi_{P1}) +
(\pi_{R2} - \pi_{P2})\,.
\eeqa
The equation for the coefficients $\alpha_{1,2}$ introduced in
\Eq{longitudinalansatz} is then obtained by substituting that 
expression in \Eq{effeqmotion1} and using \Eq{ellpiULell}, which yields
\beq
\label{longitudinalphaeq}
\sum_b (L_{ab} - \Pi_{ab})\alpha_b = 0\,.
\eeq
Therefore, the dispersion relations are obtained by solving
\beq
\label{DetLminusPi}
\mbox{Det}(L - \Pi) = 0\,.
\eeq
This yields in principle two dispersion relations with the coefficients
$\alpha_{1,2}$ obtained for each of them from \Eq{longitudinalphaeq}
and the corresponding spinors given by \Eq{longitudinalansatz}.

We obtain the antiparticle modes in similar fashion by seeking the solution
in the form
\beq
\psi^{\prime\,\mu}_L = \sum_{a = 1,2}\alpha^\prime_a U^{\prime\mu}_{La}\,,
\eeq
where
\beqa
U^{\prime\mu}_{L1} & = & \epsilon^\mu_{\ell}u_{L+}\,,\nonumber\\
U^{\prime\mu}_{L2} & = & \epsilon^\mu_{+}u_{L-}\,.
\eeqa
As in the previous case, the next step is to obtain the formulas for
$\ell_{\mu\nu\alpha}\gamma^\alpha$ and $\pi_{\mu\nu\alpha}\gamma^\alpha$
acting on the spinors $U^{\prime\mu}_{L1,2}$, the details of which
are given in the Appendix. Introducing the right-handed spinors
\beqa
U^{\prime\mu}_{R1} & = & \epsilon^\mu_{\ell}u_{R+}\,,\nonumber\\
U^{\prime\mu}_{R2} & = & \epsilon^\mu_{+}u_{R-}\,,
\eeqa
the formulas analogous to \Eq{ellpiULell} in the present case are,
\beqa
\label{ellpiUprimeLell}
\left(\ell_{\mu\nu\alpha}\gamma^\alpha\right) U^{\prime\nu}_{La} & = &
\sum_b L^\prime_{ba}U^{\prime}_{Rb\mu}\,,\nonumber\\
\left(\pi_{\mu\nu\alpha}\gamma^\alpha\right) U^{\prime\nu}_{La} & = &
\sum_b \Pi^\prime_{ba}U^{\prime}_{Rb\mu}\,,
\eeqa
where
\beqa
\label{Lprimedef}
L^\prime_{11} & = & 0\,,\nonumber\\
L^\prime_{12} = L_{21} & = & -\sqrt{2k^2}\,,\nonumber\\
L^\prime_{22} & = & (\omega - \kappa)\,,
\eeqa
and
\beqa
\label{Piprime}
\Pi^\prime_{11} & = & (\omega + \kappa)\pi_{Q1} + \pi_{Q2}\,,\nonumber\\
\Pi^\prime_{12} & = & \sqrt{-2\tilde u^2}(\pi_{R4} + \pi_{P4}) -
\sqrt{2k^2}\pi_0\,,\nonumber\\
\Pi^\prime_{21} & = & \sqrt{-2\tilde u^2}\left(\pi_{R3} + \pi_{P3}\right) -
\sqrt{2k^2}\pi_0\,,\nonumber\\
\Pi^\prime_{22} & = & (\omega - \kappa)(\pi_0 + \pi_{R1} + \pi_{P1}) +
(\pi_{R2} + \pi_{P2})\,.
\eeqa
The equation for the coefficients $\alpha^\prime_a$ is
\beq
\label{longitudinalphaprimeeq}
\sum_b (L^\prime_{ab} - \Pi^\prime_{ab})\alpha^\prime_b = 0\,,
\eeq
and in particular the dispersion relations are obtained by solving
\beq
\label{DetLminusPiprime}
\mbox{Det}(L^\prime - \Pi^\prime) = 0\,.
\eeq

\subsection{Discussion}

Besides the dispersion relation, a quantity that is physically relevant
is the proper normalization factor of the spinor solutions. In principle
such factors can be determined by mimicking the procedure followed
in the spin-1/2 case\cite{weldon:fermions,jfn1}. For the transverse mode
the analogy with the spin-1/2 case is close, but for the longitudinal
mode the treatment must take into account the fact that it involves two 
spinor solutions. In this work we have not considered the calculation
of such normalization factors.

\section{Example}
\label{sec:model}

\subsection{Model}

As a specific application of the previously developed formalism,
here we consider a background medium
composed of a spin-1/2 particle $f$ and a scalar particle $A$,
that interact with the $\lambda$ particle with the interaction Lagrangian
\beq
\label{modelLint}
L_{int} = h A \bar f_R\sigma^{\mu\nu}\partial_\mu\lambda_{L\nu} + h.c. \,.
\eeq
where the coupling parameter $h$ is inversely proportional
to some mass scale.
%
%
% fig 1
%
\begin{figure}
\begin{center}
\epsfig{file=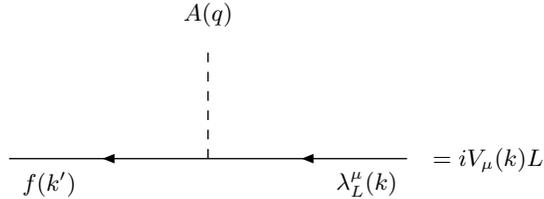,bbllx=230pt,bblly=660pt,bburx=435pt,bbury=750pt}
\end{center}
\caption[]{Vertex diagram for the spin-3/2 particle $\lambda_{L\mu}$
  with the spin-1/2 fermion $f$ and the scalar particle $A$. The
  vertex function $V_\mu(k)$ is given in \Eq{V2}.
  \label{fig:vertexdiagram}
}
\end{figure}
With the conventions specified in \Fig{fig:vertexdiagram}
this interaction gives a term $iV_\mu(k)L$ in the Feynman diagrams, where
\beq
\label{V2}
V_\mu(k) = ih\sigma_{\mu\alpha}k^\alpha \,.
\eeq
$L_{int}$ is invariant under the transformation
$\lambda_{L\mu} \rightarrow \lambda_{L\mu} + \partial_\mu\epsilon$, where
$\epsilon$ is a Dirac field, which manifests in the fact that
\beq
\label{V2transverse}
k\cdot V(k) = 0\,.
\eeq
As a consequence of this, the $\lambda$ self-energy is transverse,
as we will confirm explicitly in the calculation below.

Before entering the details of the calculation we mention the following.
The model interaction given in \Eq{modelLint} is not renormalizable.
Our attitude here is the usual one, namely that such an interaction can arise
as an effective interaction due to the exchange of heavier particles
in a more fundamental theory in which the heavier fields are integrated out.
This is analogous, for example, to the Fermi
four-fermion interaction for neutrinos. In that case the resulting
effective theory is presumed to be valid for calculating tree-level
amplitudes for external momenta much smaller than the heavy particle mass,
including the lowest order thermal loops which are just tree-level amplitudes
weighted by the appropriate thermal distributions. The results obtained this
way are valid in environments in which the thermal distributions of the heavy
particles (e.g., the $W$ gauge bisons in the neutrino case) are negligible.
In the case of neutrinos this approach leads to the
Wolfenstein formula for the neutrino index of refraction. Our
expectation is that similar considerations apply to the model
interaction of \Eq{modelLint} as well, and that the
the lessons learned by considering this example
will serve to guide the application to more general
and/or fundamental interactions of the spin-3/2 particle.

\subsection{One-loop thermal self-energy}

\begin{figure}
\begin{center}
\epsfig{file=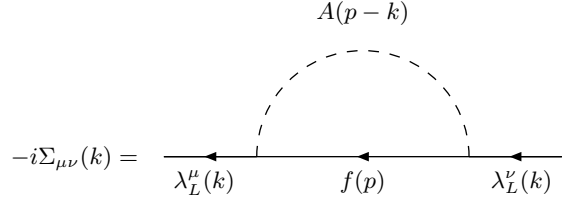,bbllx=175pt,bblly=660pt,bburx=385pt,bbury=745pt}
\end{center}
\caption[]{Diagram for the self-energy of the spin-3/2 particle $\lambda_L$
  in the background of the spin-1/2 fermion $f$ and the scalar particle $A$.
  \label{fig:selfenergydiagram}
}
\end{figure}

Our problem at hand is to compute the thermal self-energy diagram
depicted in \Fig{fig:selfenergydiagram} and then determine
the dispersion relations for the $\lambda_{L\mu}$.
%
%
% fig 1
%
In what follows we consider only the real part of the dispersion relation, 
for which we need to determine only the dispersive part of the
thermal self-energy $\Sigma_{T\mu\nu}$.
In the real-time formulation of TFT, which we will use,
the $f$ and $A$ thermal propagators as well as the self-energy
$\Sigma_{\mu\nu}$ in \Fig{fig:selfenergydiagram} are $2\times 2$ matrices.
The dispersive part of $\Sigma_{T\mu\nu}$ can be determined 
from the diagonal elements of the self-energy matrix, in particular from
the $11$ element,
\beq
\label{Sigma1}
-i\Sigma_{11\mu\nu} = \int\,\frac{d^4p}{(2\pi)^4} i\Delta_{F11}(p - k)
iR\bar V_\mu(k) iS_{F11}(p) iV_\nu(k)\,.
\eeq
In \Eq{Sigma1} the vertex function $V_\mu$ has been defined in \Eq{V2} and
$\bar V = \gamma_0 V^\dagger_\mu\gamma_0$ while the propagators in \Eq{Sigma1}
are given by
\beqa
\label{propagators11}
S_{F11}(p) & = & S_{F0}(p) + S_T(p)\,, \nonumber\\
\Delta_{F11}(p) & = & \Delta_{F0}(p) + \Delta_T(p)\,,
\eeqa
where $S_{F0}$ and $\Delta_{F0}$ stand for the vacuum propagators
\beqa
S_{F0}(p) & = & \frac{1}{\lslash{p} - m_f + i\epsilon}\,,\nonumber\\
\Delta_{F0}(p) & = & \frac{1}{p^2 - m^2_A + i\epsilon}\,,
\eeqa
and the background-dependent parts are given by
\beqa
S_T(p) & = & 2\pi i\delta(p^2 - m^2_f)\eta_f(p\cdot u)\,,\nonumber\\
\Delta_T(p) & = & -2\pi i\delta(p^2 - m^2_A)\eta_A(p\cdot u)\,,
\eeqa
with
\beqa
\eta_f(x) & = & \frac{\theta(x)}{e^{\beta x - \alpha_f} + 1} +
\frac{\theta(-x)}{e^{-\beta x + \alpha_f} + 1}\,,\nonumber\\[12pt]
\eta_A(x) & = & \frac{\theta(x)}{e^{\beta x - \alpha_A} - 1} +
\frac{\theta(-x)}{e^{-\beta x + \alpha_A} - 1}\,.
\eeqa
Here $1/\beta$ is the temperature and $\alpha_{f,A}$ are the chemical
potentials of the fermion and scalar thermal background, respectively.

The substitution of \Eq{propagators11} in
\Eq{Sigma1} gives several contributions to the self-energy. The term that
contains both $S_T$ and $\Delta_T$ contributes only to the absorptive part of
the self-energy, which we discard because we are considering only
the real part of the dispersion relation as already mentioned.
The term that contains neither $S_T$ nor $\Delta_T$ corresponds to
the pure vacuum contribution, which is not calculable in this model,
but we neglect it assuming that it is unimportant relative to the
background-dependent part. The remaining ones, that contain
either $S_T$ or $\Delta_T$, are precisely the contributions
to $\Sigma_{T\mu\nu}$ that we are after. In this way we then obtain
%
%\beq
%\Sigma_{11\mu\nu} = \Sigma_{0\mu\nu} + \Sigma_{T\mu\nu}\,,
%\eeq
%
%where $\Sigma_{0\mu\nu}$ is the pure vacuum contribution,
%
%\beq
%\label{sigma0}
%\Sigma_{0\mu\nu} = i\int\,\frac{d^4p}{(2\pi)^4} i\Delta_{F0}(p - k)
%iR\bar V_\mu(k,p) iS_{F0}(p) iV_\nu(k,p)L\,,
%\eeq
%
%
\beq
\label{sigmaT}
\Sigma_{T\mu\nu} = R\left(\Sigma^{(f)}_{\mu\nu} +
\Sigma^{(A)}_{\mu\nu}\right)L\,,
\eeq
where
\beqa
\label{sigmaTfA}
\Sigma^{(f)}_{\mu\nu} & = & -\int\,\frac{d^4p}{(2\pi)^3}
\frac{\delta(p^2 - m^2_f)\eta_f(p\cdot u)}{(p - k)^2 - m^2_A}
\bar V_\mu(k)(\lslash{p} + m_f) V_\nu(k)\,,\nonumber\\[12pt]
\Sigma^{(A)}_{\mu\nu} & = & \int\,\frac{d^4p}{(2\pi)^3}
\frac{\delta(p^2 - m^2_A)\eta_A(p\cdot u)}{(p + k)^2 - m^2_f}
\bar V_\mu(k)(\lslash{p} + \lslash{k} + m_f)V_\nu(k)\,.
\eeqa
The fact that the vertex satisfies \Eq{V2transverse}, in turn implies that
\beq
\label{kdotsigma}
k^\nu\Sigma_{T\mu\nu} = 0
\eeq
as well, as we had already anticipated.
In the following section we use the general decomposition of the self-energy
introduced in Section\ \ref{sec:generalform} and calculate the corresponding
coefficients.

\subsection{Calculation of the coefficients}

In correspondence with \Eq{sigmaT} we write
\beq
\pi_{\mu\nu\alpha} = \pi^{(f)}_{\mu\nu\alpha} + \pi^{(A)}_{\mu\nu\alpha}\,,
\eeq
where
\beq
\pi^{(f,A)}_{\mu\nu\alpha} = \frac{1}{2}\Tr\left(
\gamma_\alpha\Sigma^{(f,A)}_{T\mu\nu}\right)\,,
\eeq
with $\Sigma^{(f,A)}_{T\mu\nu}$ given in \Eq{sigmaTfA}.
The resulting formulas $\pi^{(f,A)}_{\mu\nu\alpha}$ can be written
in the form
%
%\beqa
%\pi^{(f)}_{\mu\nu\lambda} & = & -h^2 t_{\lambda\mu\rho\nu} I^{(f)\rho}\,,
%\nonumber\\
%\pi^{(A)}_{\mu\nu\lambda} & = & h^2 t_{\lambda\mu\rho\nu}
%\left[I^{(A)\rho} + k^\rho c_A\right]\,,
%\eeqa
%
\beqa
\label{piIab}
\pi^{(f)}_{\mu\nu\alpha} & = & -h^2 t_{\mu\nu\alpha\beta} I^\beta_f\,,
\nonumber\\
\pi^{(A)}_{\mu\nu\alpha} & = & h^2 t_{\mu\nu\alpha\beta}
\left[I^\beta_A + C_A k^\beta\right]\,,
\eeqa
where
%
%\beq
%\label{tdef}
%t_{\lambda\mu\rho\nu} = \frac{1}{2}k^\alpha k^\beta\mbox{Tr}\,
%L\gamma_\lambda\sigma_{\mu\alpha}\gamma_\rho\sigma_{\nu\beta}\,,
%\eeq
%
\beq
\label{tdef}
t_{\mu\nu\alpha\beta} = \frac{1}{2}k^\lambda k^\rho\mbox{Tr}\,
L\gamma_\alpha\sigma_{\mu\lambda}\gamma_\beta\sigma_{\nu\rho}\,,
\eeq
and
\beqa
C_A & = & \int\frac{d^3p}{(2\pi)^3 2E_A}\left\{
\frac{f_A(p)}{(p + k)^2 - m^2_f} + \frac{f_{\bar A}(p)}{(p - k)^2 - m^2_f}
\right\}
\nonumber\\
I^\mu_A & = & \int\frac{d^3p}{(2\pi)^3 2E_A}\, p^\mu \left\{
\frac{f_A(p)}{(p + k)^2 - m^2_f} - \frac{f_{\bar A}(p)}{(p - k)^2 - m^2_f}
\right\}
\nonumber\\
I^\mu_f & = & \int\frac{d^3p}{(2\pi)^3 2E_f}\, p^\mu \left\{
\frac{f_f(p)}{(p - k)^2 - m^2_A} - \frac{f_{\bar f}(p)}{(p + k)^2 - m^2_A}
\right\} \,.
\eeqa
In these integrals $f_{A,f}(p)$ are the thermal distribution functions
\beqa
f_A(p) & = & \frac{1}{e^{\beta E_A - \alpha_A} - 1}\,,\nonumber\\ 
f_f(p) & = & \frac{1}{e^{\beta E_f - \alpha_f} + 1} \,,
\eeqa
and the anti-particle counterparts $f_{\bar A,\bar f}$ are obtained from them
by making replacements $\alpha_{A,f}\rightarrow -\alpha_{A,f}$.

The integrals $I^\mu_{A,f}$ can be expressed in the form
\beq
\label{Iab}
I^\mu_X = A_X k^\mu + B_X u^\mu \qquad (X = f,A)\,,
\eeq
where the coefficients $A_{A,f}, B_{A,f}$ can be expressed in terms
of scalar integrals by inverting the equations
\beqa
k\cdot I_X & = & k^2 A_X + \omega B_X\,,\nonumber\\
u\cdot I_X & = & \omega A_X + B_X\,,
\eeqa
implied by \Eq{Iab}. This procedure yields the formulas
\beqa
A_X & = & \frac{1}{\kappa} L^{(1)}_X\,,\nonumber\\
B_X & = & L^{(2)}_X - \frac{\omega}{\kappa} L^{(1)}_X\,,
\eeqa
where
\beqa
\label{L12fermion}
L^{(1)}_f & = & \int\frac{d^3p}{(2\pi)^3 2E_f}\, \hat k\cdot\vec p \left\{
\frac{f_f(p)}{(p - k)^2 - m^2_A} - \frac{f_{\bar f}(p)}{(p + k)^2 - m^2_A}
\right\}\,,\nonumber\\
L^{(2)}_f & = & \frac{1}{2}\int\frac{d^3p}{(2\pi)^3}\,\left\{
\frac{f_f(p)}{(p - k)^2 - m^2_A} - \frac{f_{\bar f}(p)}{(p + k)^2 - m^2_A}
\right\}\,,
\eeqa
while the analogous formulas for $L^{(1,2)}_A$
are obtained from \Eq{L12fermion}
by making the replacements $k\rightarrow -k$,
$f_{f,\bar f}\rightarrow f_{A,\bar A}$ and $m_A\rightarrow m_f$
on the right-hand side.

Substituting \Eq{Iab} in \Eq{piIab}, $\pi_{\mu\nu\alpha}$ is then given by
%
%\beq
%\label{pioneloop}
%\pi_{\mu\nu\lambda} =
%h^2(a_A + c_A - a_f)t_{\lambda\mu\rho\nu} k^\rho + 
%h^2(b_A - b_f) t_{\lambda\mu\rho\nu} u^\rho\,.
%\eeq
%
\beq
\label{pioneloop}
\pi_{\mu\nu\alpha} =
h^2(C_A + A_A - A_f)t_{\mu\nu\alpha\beta} k^\beta + 
h^2(B_A - B_f) t_{\mu\nu\alpha\beta} u^\beta\,.
\eeq
Straightforward evaluation of the trace in \Eq{tdef} yields
%
%\beq
%\label{tresult}
%t_{\lambda\mu\rho\nu} = -k^2\gtilde_{\mu\lambda}\gtilde_{\nu\rho} -
%k^2\gtilde_{\mu\rho}\gtilde_{\nu\lambda} +
%k^2\gtilde_{\mu\nu}\gtilde_{\lambda\rho} - k_\lambda k_\rho\gtilde_{\mu\nu} -
%i\left[k_\lambda\epsilon_{\mu\nu\rho\alpha}k^\alpha + 
%k_\rho\epsilon_{\mu\nu\lambda\alpha}k^\alpha\right]\,,
%\eeq
%
\beq
\label{tresult}
t_{\mu\nu\alpha\beta} = -k^2\gtilde_{\mu\alpha}\gtilde_{\nu\beta} -
k^2\gtilde_{\mu\beta}\gtilde_{\alpha\nu} +
k^2\gtilde_{\mu\nu}\gtilde_{\alpha\beta} - k_\alpha k_\beta\gtilde_{\mu\nu} -
i\left[k_\alpha\epsilon_{\mu\nu\beta\rho}k^\rho + 
k_\beta\epsilon_{\mu\nu\alpha\rho}k^\rho\right]\,.
\eeq
%
%from which we obtain
%
%\beqa
%\label{tgamma}
%t_{\lambda\mu\rho\nu} k^\rho & = & -k^2\left[
%k_\lambda \gtilde_{\mu\nu} + i\epsilon_{\mu\nu\lambda\alpha}k^\alpha\right]
%\,,\nonumber\\
%t_{\lambda\mu\rho\nu} u^\rho & = & k^2\left[
%\tilde u_\lambda\gtilde_{\mu\nu} -
%\tilde u_\mu\gtilde_{\lambda\nu} -
%\tilde u_\nu\gtilde_{\mu\lambda}\right] -
%(k\cdot u)k_\lambda\gtilde_{\mu\nu}\nonumber\\
%&&\mbox{} -
%ik_\lambda\epsilon_{\mu\nu\alpha\beta} u^\alpha k^\beta -
%i(k\cdot u)\epsilon_{\mu\nu\lambda\alpha} k^\alpha\,.
%\eeqa
%
%\beqa
%\label{tgamma}
%t_{\mu\nu\alpha\beta} k^\beta & = & -k^2\left[
%k_\alpha \gtilde_{\mu\nu} + i\epsilon_{\mu\nu\alpha\beta}k^\beta\right]
%\,,\nonumber\\
%t_{\mu\nu\alpha\beta} u^\beta & = & k^2\left[
%\tilde u_\alpha\gtilde_{\mu\nu} -
%\tilde u_\mu\gtilde_{\alpha\nu} -
%\tilde u_\nu\gtilde_{\mu\alpha}\right] -
%(k\cdot u)k_\alpha\gtilde_{\mu\nu}\nonumber\\
%&&\mbox{} -
%ik_\alpha\epsilon_{\mu\nu\lambda\rho} u^\lambda k^\rho -
%i(k\cdot u)\epsilon_{\mu\nu\alpha\beta} k^\beta\,.
%\eeqa
%
Rewriting this expression in terms of $R_{\mu\nu}, Q_{\mu\nu}$ and
$\ell_{\mu\nu\alpha}$ by means of \Eqs{elldef}{RQP}, and then
substituting the result in \Eq{pioneloop} we obtain the one-loop
result for $\pi_{\mu\nu\alpha}$ in the form given in
\Eqs{piprimedef}{piparametrization}, with the coefficients 
\beqa
\label{pieffgenpiABC}
\pi_{0} & = & -h^2 k^2(C_A + A_A - A_f) - h^2\omega(B_A - B_f)\,,\nonumber\\
\pi_{R1} & = & -h^2 k^2(C_A + A_A - A_f) - 2h^2\omega(B_A - B_f)\,,\nonumber\\
\pi_{Q1} & = & -h^2 k^2(C_A + A_A - A_f)\,,\nonumber\\
\pi_{P1} & = & h^2 \kappa(B_A - B_f)\,,\nonumber\\
\pi_{R3} & = & \pi_{R4} = \pi_{Q2} = -\pi_{R2} =
-h^2 k^2(B_A - B_f)\,,\nonumber\\
\pi_{P2} & = & \pi_{P3} = \pi_{P4} = 0\,.
\eeqa

\subsection{Discussion}

As a specific example let us consider the situation in which the mass
of the scalar boson is much greater than the other relevant energy scales,
i.e.,
\beq
\label{backgroundconditions}
T, \mu, \omega, \kappa, m_f \ll m_A\,,
\eeq
so that, in particular, there are no $A$ scalars in the background.
Thus in this case,
\beq
C_A = L^{(1)}_f = L^{(1,2)}_A \simeq 0\,,
\eeq
while
\beq
L^{(2)}_f = \frac{-1}{4 m^2_A}(n_f - n_{\bar f})\,,
\eeq
and therefore,
\beqa
\label{BABf}
B_A & = & A_f = A_A \simeq 0\,,\nonumber\\
B_f & = & \frac{-1}{4 m^2_A}(n_f - n_{\bar f})\,.
\eeqa
From \Eq{pieffgenpiABC},
\beqa
\label{pieffgenpiABCmodel}
\pi_0 & = & \frac{1}{2}\pi_{R1} = h^2B_f\omega\,,\nonumber\\
\pi_{R3} & = & \pi_{R4} = \pi_{Q2} = -\pi_{R2} = h^2 B_f k^2\,,\nonumber\\
\pi_{P1} & = & -h^2 B_f \kappa\,,\nonumber\\
\pi_{Q1} & = & \pi_{P2} = \pi_{P3} = \pi_{P4} = 0\,,
\eeqa
from which we can now determine the dispersion relations.

\subsubsection{Transverse mode}

For the transverse modes, the parameters $a_{\pm}, b_{\pm}$ defined
in \Eq{abpm} are then
\beqa
\label{abmodel}
a_{\pm} & = & -h^2 B_f (\omega \mp \kappa)\,,\nonumber\\
b_{\pm} & = & h^2 B_f k^2\,.
\eeqa
Substituting \Eq{abmodel} into the dispersion relation equation for the 
transverse modes, \Eqs{transdisprel1}{transdisprel1neg}, the $B_f$
term cancels in both cases and the solutions are
\beqa
\omega & = & \kappa\,,\nonumber\\
\bar\omega & = & \kappa\,,
\eeqa
for the particle and antiparticle, respectively.
Therefore the dispersion relation for the transverse mode
is not modified in the presence of the background. However,
it should be remembered that this result holds under the conditions stated in
\Eq{backgroundconditions}, and for other  conditions and/or regimes
(e.g., $\omega,\kappa \gg m_{A,f})$ the dispersion
relation is in general modified. In the spirit of our presentation
of this model being for illustrative purposes, here we do not pursue
this further and turn instead to the longitudinal mode.

\subsubsection{Longitudinal mode}

For the longitudinal mode, using \Eq{Piprime}
the matrix $\Pi$ defined in \Eq{Pi} is given by
\beq
\label{Pimatrixmodel}
\Pi = h^2 B_f(\omega + \kappa)\left(
\begin{array}{cc}
\omega - \kappa & \qquad \sqrt{2k^2}\\
\sqrt{2k^2} & \qquad 2(\omega + \kappa)
\end{array}
\right)\,.
\eeq
Using \Eqs{Ldef}{Pimatrixmodel}, the equation [\Eq{DetLminusPi}]
for the longitudinal dispersion relation is
\beq
-k^2 x(1 - 2x) -2k^2(1 - x)^2 = 0\,,
\eeq
where we have defined
\beq
x = h^2 B_f(\omega + \kappa) \,.
\eeq
The solution with $k^2 \not= 0$ is $x = 2/3$, which gives the dispersion
relation
\beq
\label{longdisprelmodel}
\omega_\kappa = M - \kappa\,,
\eeq
where
\beq
\label{Mdef}
M \equiv \frac{2}{3h^2 B_f}\,.
\eeq
For completeness we note that going back to
\Eqs{longitudinalansatz}{longitudinalphaeq}, the corresponding spinor is
\beq
\psi^\mu_{L\ell} = \alpha_1\, \epsilon^\mu_{\ell} u_{L-} +
\alpha_2\, \epsilon^\mu_{-}u_{L+}\,,
\eeq
where, up to a normalization factor,
\beqa
\alpha_1 & = & \frac{1}{\sqrt{2}}\,,\nonumber\\
\alpha_2 & = & \sqrt{1 - \frac{2\kappa}{M}}\,.
\eeqa
In similar fashion, from \Eq{Piprime},
\beq
\label{Pimatrixmodelprime}
\Pi^\prime = h^2 B_f(\omega - \kappa)\left(
\begin{array}{cc}
\omega + \kappa & \qquad -\sqrt{2k^2}\\
-\sqrt{2k^2} & \qquad 2(\omega - \kappa)
\end{array}
\right)\,.
\eeq
\Eqs{longitudinalphaprimeeq}{DetLminusPiprime} have the solution
\beq
\omega^\prime_\kappa = M + \kappa\,,
\eeq
with the corresponding spinor
\beq
\psi^{\prime\,\mu}_{L\ell} =
\alpha^\prime_1\epsilon^\mu_{\ell}u_{L+} +
\alpha^\prime_2\epsilon^\mu_{+}u_{L-}\,,
\eeq
where again, up to a normalization factor
\beqa
\alpha^\prime_1 & = & \frac{1}{\sqrt{2}}\,,\nonumber\\
\alpha^\prime_2 & = & -\sqrt{1 + \frac{2\kappa}{M}}\,.
\eeqa

The proper interpretation of these branches in terms of particle-antiparticle
or hole-antihole modes depends on the sign of $M$ and the relative
size of $\kappa$ and $M$. For a specific example, suppose that the
medium is such that $n_f > n_{\bar f}$. From \Eqs{BABf}{Mdef}
it follows that $M < 0$. Then for $\kappa > |M|$ the dispersion relations
\beqa
\omega^\prime_\kappa & = & \kappa - |M|\,,\nonumber\\
\omega_\kappa & = & -\kappa - |M| \equiv -\bar\omega_\kappa\,,
\eeqa
resemble the dispersion relations for the neutrino ($\omega^\prime_\kappa$)
and antineutrino ($\bar\omega_\kappa$) in an
electron background\cite{palpham,jfn1,dnt}.

Finally we mention the following.
Consistency with the conditions stated in \Eq{backgroundconditions}
imposes some requirements for this solution to be valid,
but it is easy to see that they can be satisfied in this model example.
For instance, remembering that $h$ is inversely proportional
to some mass scale, suppose that $h \sim 1/m_f$. The above result then gives
$\omega_\kappa \sim O(m^2_A m^2_f/n_f)$. Assuming that $T \gg m_f$,
so that $n_f \sim T^3$, \Eq{backgroundconditions} is satisfied for
\beq
(m_A m^2_f)^{1/3} \ll T \ll m_A \,.
\eeq

It has not been our intention in this section to study the model
exhaustively. Rather, our purpose in going through these details has
been to show that the solutions given above give
results that are analogous to other well studied systems and
consistent with the assumptions that we have made in the model.

\section{Conclusions}

In the present work we have used the methods of TFT to treat
the propagation of a chiral spin-3/2 particle $\lambda^\mu_{L}$
in a background medium using the methods of TFT, in analogy with the familiar
cases of a photon or a neutrino propagating in a matter background.
The essential ingredient of the method is the general decomposition of the
self-energy in terms of a set of scalar functions, each one corresponding
to an independent tensor constructed using the momentum vectors available in
the system (the momentum of the particle $k^\mu$ and the background velocity
four-vector $u^\mu$), as well as the gamma matrices, the metric tensor
and the Levi-Civita tensor. 
Throughout this work we have assumed that the interactions
of the $\lambda_L$ particle with the thermal background particles are such
that the thermal self-energy is transverse to $k^\mu$.
On the basis of such decomposition we showed that there is a transverse mode
in which the spin-3/2 spinor is transverse to $k$, and
two other modes that involve the longitudinal polarization vector.
In each case we obtained the equation for the dispersion relation
in terms of the self-energy scalar functions, as well as the corresponding
spin-3/2 spinor. Finally, we illustrated the application of the formalism
by computing the 1-loop TFT expression for the self-energy in a model in
which the $\lambda^\mu_L$ propagates in a thermal background composed
of spin-1/2 fermion and a scalar particle, and applying the general
results to determine the dispersion relations and corresponding spinors.
As we showed, the results so obtained share some
resemblance and analogies with the photon and the chiral spin-1/2 fermion
case, but there are some differences as well.

The present work provides the groundwork for considering problems
involving a chiral spin-3/2 particle propagating in a medium,
which can be relevant in cosmological or astrophysical
contexts of current research interest.
This work also opens the path to consider the effects of the medium
on the electromagnetic properties of a massless chiral spin-3/2 particle,
in analogy with the neutrino case\footnote{%
For a review of the electromagnetic couplings of a neutrino in a background
medium see for example \Ref{raffelt:book}}.
Such induced electromagnetic couplings can have effects in the
cosmological or astrophysical contexts in which the electromagnetic
couplings are involved.

\begin{acknowledgments}
The work of SS is partially supported by DGAPA-UNAM (Mexico) project
No. IN110815.
\end{acknowledgments}

\appendix

\section{Formulas for the longitudinal spinor}
\label{sec:appendixA}

\subsection{Derivation of \Eqs{Ldef}{Pi}}

From the equations satisfied by the spinors $U^\mu_{L1,2}$ defined in \Eq{UL12}
we derive the formulas for $\ell_{\mu\nu\alpha}\gamma^\alpha U^\nu_{L1,2}$
and $\pi_{\mu\nu\alpha}\gamma^\alpha U^\nu_{L1,2}$. We consider each of them
one by one.

Consider $\epsilon^\mu_{\ell} u_{L-}$ first.
From the definition of $\ell_{\mu\nu\alpha}, P_{\mu\nu}$ and
$\epsilon^\mu_\ell$ in \Eqss{elldef}{RQP}{epsilonell},
\beq
\label{ellepsilonellPrelation}
\ell_{\mu\nu\alpha}\epsilon^\nu_{\ell} = \sqrt{k^2}P_{\mu\alpha} \,,
\eeq
and using \Eq{RPgammauL},
\beq
\label{ellUL1}
\left(\ell_{\mu\nu\alpha}\gamma^\alpha\right)\epsilon^\nu_{\ell} u_{L-} = 
\sqrt{2k^2}g_{\mu\nu}\epsilon^\nu_{-} u_{R+} \,.
\eeq
Using the orthonormality relations of $\epsilon^\mu_\ell$
[e.g., \Eq{RQPepsilon}],
\beq
\label{piprimeepsilonell}
\pi^\prime_{\mu\nu\alpha}\epsilon^\nu_{\ell} =
\pi_{Q1}\epsilon_{\ell\mu}k_\alpha +
\pi_{Q2}\epsilon_{\ell\mu}u_\alpha -
\sqrt{-\tilde u^2}\pi_{R3}R_{\mu\alpha} -
\sqrt{-\tilde u^2}\pi_{P3}P_{\mu\alpha}\,,
\eeq
and from \Eqs{slashuLrelations}{RPgammauL}
\beq
\label{piprimeUL1}
(\pi^\prime_{\mu\nu\alpha}\gamma^\alpha)\epsilon^\nu_{\ell}u_{L-} =
\left\{
\left[(\omega - \kappa)\pi_{Q1} + \pi_{Q2}\right]\epsilon_{\ell\mu}u_{R-} +
\sqrt{-2\tilde u^2}\left(\pi_{R3} - \pi_{P3}\right)\epsilon_{-\mu}u_{R+}
\right\}\,.
\eeq
For $\epsilon^\mu_{-}u_{L+}$, using \Eq{epsilonidentitiesplusminus},
\beq
\ell_{\mu\nu\alpha}\gamma^\alpha\epsilon^\nu_{-} =
\frac{-\kappa}{\sqrt{-\tilde u^2}}
\left(\epsilon_{-\mu}\epsilon_\ell\cdot\gamma -
\epsilon_{\ell\mu}\epsilon_{-}\cdot\gamma\right)\,,
\eeq
and from \Eqs{slashuLrelations}{epsilondotgammauL}
\beq
\label{ellUL2}
(\ell_{\mu\nu\alpha}\gamma^\alpha)\epsilon^\nu_{-}u_{L+} =
\left\{(\omega + \kappa)\epsilon_{-\mu} u_{R+} +
\sqrt{2k^2}\epsilon_{\ell\mu}u_{R-}\right\}\,.
\eeq
In similar fashion, using
\Eqsss{transversality}{RQPepsilon}{slashuLrelations}{epsilondotgammauL}
\beqa
\label{piprimeUL2}
(\pi^\prime_{\mu\nu\alpha}\gamma^\alpha)\epsilon^\nu_{-}u_{L+} & = & 
\left\{
\left[(\omega + \kappa)(\pi_{R1} - \pi_{P1}) + (\pi_{R2} - \pi_{P2})
\right]\epsilon_{-\mu}u_{R+}\right.\nonumber\\
&&\mbox{} +
\left.\sqrt{-2\tilde u^2}(\pi_{R4} - \pi_{P4})\epsilon_{\ell\mu}u_{R-}
\right\}\,.
\eeqa
The results given in \Eqsss{ellUL1}{piprimeUL1}{ellUL2}{piprimeUL2}
are summarized in \Eq{ellpiULell}.

\subsection{Derivation of \Eqs{Lprimedef}{Piprime}}

In similar fashion we derive the formulas for
$\ell_{\mu\nu\alpha}\gamma^\alpha U^{\prime\,\nu}_{L1,2}$ and
$\pi_{\mu\nu\alpha}\gamma^\alpha U^{\prime\,\nu}_{L1,2}$.
Thus, using \Eqs{RPgammauL}{ellepsilonellPrelation},
\beq
(\ell_{\mu\nu\alpha}\gamma^\alpha)\epsilon^\nu_{\ell} u_{L+} =
-\sqrt{2k^2}\epsilon_{+\mu}u_{R-}\,,
\eeq
and using
\Eqss{epsilonidentitiesplusminus}{slashuLrelations}{epsilondotgammauL}
\beq
(\ell_{\mu\nu\alpha}\gamma^\alpha)\epsilon^\nu_{+} u_{L-} =
(\omega - \kappa)\epsilon_{+\mu}u_{R-} - \sqrt{2k^2}\epsilon_{\ell\mu}u_{R+}\,.
\eeq
From \Eq{piprimeepsilonell}, and using \Eqs{slashuLrelations}{RPgammauL}
\beq
\label{piprimeUprimeL1}
(\pi^\prime_{\mu\nu\alpha}\gamma^\alpha)\epsilon^\nu_{\ell}u_{L+} =
\left\{
\left[(\omega + \kappa)\pi_{Q1} + \pi_{Q2}\right]\epsilon_{\ell\mu}u_{R+} +
\sqrt{-2\tilde u^2}\left(\pi_{R3} + \pi_{P3}\right)\epsilon_{+\mu}u_{R-}
\right\}\,,
\eeq
and finally from
\Eqsss{transversality}{RQPepsilon}{slashuLrelations}{epsilondotgammauL},
\beqa
\label{piprimeUprimeL2}
(\pi^\prime_{\mu\nu\alpha}\gamma^\alpha)\epsilon^\nu_{+}u_{L-} & = & 
\left\{
\left[(\omega - \kappa)(\pi_{R1} + \pi_{P1}) + (\pi_{R2} + \pi_{P2})
\right]\epsilon_{+\mu}u_{R-}\right.\nonumber\\
&&\mbox{} +
\left.\sqrt{-2\tilde u^2}(\pi_{R4} + \pi_{P4})\epsilon_{\ell\mu}u_{R+}
\right\}\,.
\eeqa
The above results are summarized in \Eq{ellpiUprimeLell}.

\end{document}